\documentclass[aps, superscriptaddress, letterpaper, twocolumn, notitlepage, 10pt]{revtex4-2}
\usepackage{graphicx}
\usepackage{amsmath,amssymb}
\usepackage{newtxtext}
\usepackage{newtxmath}
\usepackage{verbatim}
\usepackage{xfrac}
\usepackage[utf8]{inputenc}
\usepackage[x11names]{xcolor}
\usepackage[unicode,colorlinks,citecolor=Blue3,linkcolor=Blue3,bookmarks=true]{hyperref}

\newcommand{\eg}{{\it e.g.,\,}}
\newcommand{\intinfty}{\displaystyle\int_{-\infty}^{\infty}}
\newcommand{\Tr}{\mathop{\rm Tr}\nolimits}
\newcommand{\bra}[1]{\langle{#1}|}
\newcommand{\ket}[1]{|{#1}\rangle}
\newcommand{\mean}[1]{\langle{#1}\rangle}

\begin{document}


\title{Robustness of negativity of the Wigner function to dissipation}

\author{B.\,Nugmanov}\thanks{The first two authors contributed equally to this work}

\affiliation{Russian Quantum Center, Skolkovo IC, Bolshoy Bulvar 30, bld. 1, Moscow, 121205, Russia}

\author{N.\,Zunikov}\thanks{The first two authors contributed equally to this work}

\affiliation{Russian Quantum Center, Skolkovo IC, Bolshoy Bulvar 30, bld. 1, Moscow, 121205, Russia}

\author{F.\,Ya.Khalili}

\affiliation{Russian Quantum Center, Skolkovo IC, Bolshoy Bulvar 30, bld. 1, Moscow, 121205, Russia}

\affiliation{NTI Center for Quantum Communications, National University of Science and Technology MISiS, Leninsky prospekt 4, Moscow 119049, Russia}

\date{\today}

\begin{abstract}
  Non-Gaussian quantum states, described by negative valued Wigner functions, are important both for fundamental tests of quantum physics and for emerging quantum information technologies. However, they are vulnerable to dissipation. It is known, that the Wigner functions negativity could exist only if the overall quantum efficiency $\eta$ of the setup is higher than 1/2. Here we prove that this condition is not only necessary but also a sufficient one: the negativity always persists while this condition is fulfilled. At the same time, in the case of bright (multi-photon) non-Gaussian quantum states, the negativity dependence on $\eta$ is highly non-linear. With the loss of several photons, it drops by orders of magnitude, hampering its experimental detection.
\end{abstract}

\maketitle


\paragraph{Introduction.}

Quantum states of continuous-variable (CV) systems \cite{Braunstein_RMP_77_513_2005} can be conveniently described in terms of quantum quasi-probability distributions (QPDs) \cite{Cahill_PR_177_1882_1969, Schleich2001}, like the Husimi $Q$-function \cite{Husimi_PPMSJ3_22-264_1940}, the Wigner $W$-function \cite{Wigner_PR_40_749_1932}, and the Glauber-Sudarshan $P$-function \cite{Glauber_PR_131_2766_1963}. These functions, defined in the phase space, resemble the classical position/momentum probability distributions, while having one-to-one correspondence with the density operators of the respective quantum states. They can be reconstructed experimentally using the quantum tomography procedure \cite{Vogel_PRA_40_2847_1989}.

In the case of quantum states with Gaussian shape of the Wigner function --- the {\it Gaussian} quantum states --- this shape exactly coincides with corresponding classical position/momentum probability distribution. At the same time, according to the Hudson theorem \cite{Hudson_RMP_6_249_1974}, the Wigner functions of all other pure quantum states --- the {\it non-Gaussian} ones --- take negative values, which is evidently impossible for classical probability distributions.

Several interrelated features, important for both the fundamental test of applicability of quantum physics to macroscopic objects \cite{Marshall_PRL_91_130401_2003, Romero-Isart_NJP_12_033015_2010, 10a1KhDaMiMuYaCh} and to the emerging quantum information technologies \cite{Lvovsky_2006_16985, Walschaers_PRXQ_2_030204_2021}, make the non-Gaussian states ``more quantum'' than the Gaussian ones. The Gaussian states never can be orthogonal to each other, while, evidently, the orthogonality is of crucial importance for many quantum phenomena. The Gaussian states allow for a local hidden-variable description in terms of position and momentum; therefore, all experiments which rely on the Gaussian states and linear (position/momentum) measurements, could be explained within the local hidden-variables paradigm \cite{Bell_1987}. The non-Gaussian states are required for protocols that cannot be efficiently simulated by a classical computer \cite{Mari_PRL_109_230503_2012}.

At the same time, it is well known that the non-Gaussian states are vulnerable to dissipation, which converts them into non-coherent mixes of Gaussian ones. This feature is a fundamental one and can be traced up to the central limit theorem and Zurek's environment-induced superselection principle \cite{Zurek_RMP_75_715_2003}. It was shown in Ref.\,\cite{Leonhardt_PRL_72_4086_1994}, that the inequality
\begin{equation}\label{one-half}
  \eta > \frac{1}{2}
\end{equation}
where $\eta$ is the overall quantum efficiency of the setup (including, \eg the detection quantum efficiency) is the {\it necessary} condition for the Wigner function's negativity. But the natural question arises, is this condition also a {\it sufficient} one? It is reasonable to expect, that some pure non-Gaussian states, in particular bright (multi-photon) ones, are more fragile than other ones and lose their negativity at bigger values of $\eta$ than $1/2$. The broadly accepted rule is that the negativity vanishes with the loss of a few photons.

In this paper we show that the conditions \eqref{one-half} is not only necessary but also a  sufficient one: while it is fulfilled, the negativity of pure non-Gaussian states persists. At the same time, we show also that in the case of bright quantum states, the negativity dependence on $\eta$ is highly non-linear. After the loss of several photons, it drops by orders of magnitude, which could make its experimental detection problematic.

\paragraph{Quasi-probability distributions.}

In this section, we review the main features of the QPDs, known in literature, and introduce the main notations, used in this paper.

Let $\hat{a}$, $\hat{a}^\dag$ be the annihilation and creation operator a harmonic oscillator (\eg a mode of electromagnetic field), satisfying the standart commutation relation $[\hat{a}, \hat{a}^\dag] = 1$. Introduce the s-parametrized set of characteristic functions as follows \cite{Cahill_PR_177_1882_1969, Braunstein_RMP_77_513_2005}:
\begin{equation}\label{C_s}
  C(z,s) = \Tr\bigl[\hat{\rho}e^{i(z^*\hat{a} + z\hat{a}^\dag)}\bigr]e^{s|z|^2/2} \,,
\end{equation}
where $\rho$ is the density matrix, $s$ is a real parameter with $|s|\le1$, and $z$ a complex number. The inverse Fourier transform of $C(z,s)$ produces the corresponding s-parametrized set of the QPDs:
\begin{equation}
  W(\alpha,s) = \intinfty C(z,s)e^{-i(z^*\alpha + z\alpha^*)}\,\frac{d^2z}{\pi^2} \,,
\end{equation}
where $\alpha = \frac{q+ip}{\sqrt{2}}$. The important special cases of $W(\alpha,s)$ are the Husimi $Q$-function, the Wigner function $W$, and the Glauber's $P$-distibution:
\begin{equation}
  W(\alpha,-1) = Q(\alpha) \,, \quad
  W(\alpha,0) = W(\alpha) \,, \quad
  W(\alpha,1) = P(\alpha) \,.
\end{equation}
They can be expressed through each other using the following relation:
\begin{equation}\label{C_shift}
  C(z,s') = C(z,s)e^{(s'-s)|z|^2/2} .
\end{equation}

\begin{figure}
   \includegraphics{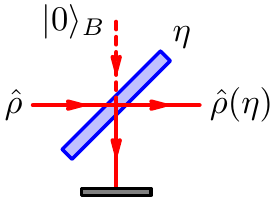}
  \caption{Model of the dissipation; $\eta$: the overall quantum efficiency.}\label{fig:losses}
\end{figure}

Consider now the dissipation. Following approach of Ref.\,\cite{Leonhardt_PRL_72_4086_1994}, we model it by means of an imaginary beamsplitter with the power transmissivity $\eta$,
which couples the considered mode with the effective heat bath mode, which we assume to be in the ground state $\ket{0_B}$, see Fig.\,\ref{fig:losses}. The density operator of the resulting quantum state is equal to
\begin{equation}\label{rho_eta}
  \hat{\rho}(\eta) = \Tr_B\bigl(
      \hat{\mathcal{U}}\,\hat{\rho}\otimes\ket{0_B}\bra{0_B}\,\hat{\mathcal{U}}^\dagger
    \bigr) ,
\end{equation}
where $\Tr_B$ is the partial trace taken over the heat bath subspace and the unitary operator $\hat{\mathcal{U}}$ describes the beamsplitter action:
\begin{equation}\label{calU_d}
  \hat{\mathcal{U}}^\dagger\hat{a}\hat{\mathcal{U}}
  = \sqrt{\eta}\,\hat{a} + \sqrt{1-\eta}\,\hat{a}_B \,,
\end{equation}
where $\hat{a}_B$ is the annihilation operator of the heat bath mode.

It follows from Eqs.\,(\ref{C_s}, \ref{rho_eta}, \ref{calU_d}), that the resulting characteristic function is equal to
\begin{multline}\label{C_eta_1}
  C(z,s,\eta) = \Tr\bigl[\hat{\rho}(\eta)e^{i(z^*\hat{a} + z\hat{a}^\dag)}\bigr]e^{s|z|^2/2}\\
  =  C(\sqrt{\eta}z,s)\exp\biggl[-\frac{(1-s)(1-\eta)}{2}|z|^2\biggr] ,
\end{multline}
or, alternatively,
\begin{equation}\label{C_eta_2}
  C(z,s,\eta) = C(z',s') \,.
\end{equation}
Here
\begin{equation}\label{zs_prime}
  z' = \sqrt{\eta}z \,, \quad  s' = \frac{s + \eta - 1}{\eta}
\end{equation}
are, respectively, the reduced by dissipation of value $z$ and the new effective value of the parameter $s$.

Inverse Fourier transform of Eq.\,\eqref{C_eta_1} gives the corresponding QPD after the losses:
\begin{multline}\label{W_eta}
  W(\alpha,s,\eta) = \intinfty C(z,s,\eta)e^{-i(z^*\alpha + z\alpha^*)}\,\frac{d^2z}{\pi^2}\\
  = \intinfty W(\beta,s)B(\alpha - \sqrt{\eta}\beta,s,\eta)d^2\beta \,,
\end{multline}
where
\begin{equation}
  B(\alpha,s,\eta) = \frac{2}{\pi(1-s)(1-\eta)}
    \exp\biggl[-\frac{2|\alpha|^2}{(1-s)(1-\eta)}\biggr]
\end{equation}
is the Gaussian blurring kernel.

\paragraph{Sufficient condition for the Wigner function negativity.}

It follows from Eqs.\,(\ref{C_eta_2}, \ref{zs_prime}) that if $s<1$, then the losses transform the initial QPD into the more smooth one with $s'<s$. In particular, is $s=0$ (the Wigner function) and $\eta=1/2$, then $s'=-1$, which corresponds to the non-negative everywhere Husimi function. So indeed \eqref{one-half} is the necessary condition for the Wigner function's negativity \cite{Leonhardt_PRL_72_4086_1994}.

Now we prove the sufficiency of this condition. Let $W_{\ket{\psi}}(\alpha,s)$ be the QPD of some pure non-Gaussian quantum state $\ket{\psi}$ and $s_0$ be the maximal value of $s$ for which $W_{\ket{\psi}}(\alpha,s)$ is non-negative everywhere. Evidently, $s_0\ge-1$ because the Husimi function $W(\alpha,-1)$ of any quantum state is non-negative everywhere, and $s_0<0$ because our quantum state is a non-Gaussian one and therefore its Wigner function $W_{\ket{\psi}}(\alpha,0)$ takes negative values. Let also $W_{\ket{\beta}}(\alpha,s)$ be the QPD of a coherent state $\ket{\beta}$.

We use the proof by contradiction. Suppose that $s_0>-1$ and consider the scalar product
\begin{equation}\label{prod_s}
  (W_{\ket{\psi}},W_{\ket{\beta}})
  = \intinfty W_{\ket{\psi}}(\alpha,s_0)W_{\ket{\beta}}(\alpha,-s_0)d^2\alpha \,.
\end{equation}
Note that $W_{\ket{\psi}}(\alpha,s_0)\ge0$ due to the definition of $s_0$ and $W_{\ket{\beta}}(\alpha,-s_0)>0$ because it is a Gaussian function. Therefore,
\begin{equation}\label{prod_positive}
  (W_{\ket{\psi}}, W_{\ket{\beta}}) > 0 \,.
\end{equation}
At the same time, it is easy to show that \eqref{prod_s} is invariant with respect to $s_0$. Therefore,
\begin{equation}\label{prod_0}
  (W_{\ket{\psi}},W_{\ket{\beta}})
  = \intinfty W_{\ket{\psi}}(\alpha,0)W_{\ket{\beta}}(\alpha,0)d^2\alpha > 0
\end{equation}
for all coherent states $\ket{\beta}$. According to the reasoning of Hudson \cite{Hudson_RMP_6_249_1974}, this means that $W_{\ket{\psi}}(\alpha,0)$ is a Gaussian function. So, we came to the contradiction.

However, if $s_0=-1$, then $W_{\ket{\beta}}(\alpha,-s_0)=\delta^2(\alpha-\beta)$, where $\delta^2$ is the two-dimensional $\delta$-function, and the strict inequality \eqref{prod_positive} relaxes to $(W_{\ket{\psi}},W_{\ket{\beta}}) \ge 0$. In this cases, no contradiction arises. Therefore, for all {\it pure non-Gaussian states}, $s_0=-1$, which means that the condition \eqref{one-half} is the {\it sufficient} one for preserving the Wigner function's negativity.

It have to be emphasized, that this is not the case for {\it mixed} quantum states. For example, in the case of the mix $\hat{\rho}=\frac{1}{2}\left[\ket{0}\bra{0} + \ket{1}\bra{1} + \frac{1}{2}\left(\ket{0}\bra{1} + \ket{1}\bra{0}\right)\right]$, negativity exists only if $\eta>7/8$.

\paragraph{Rate of the negativity degradation.}

Here we explore dependence of the Wigner function negativity on the factor $\eta$. Several quantitative measure of the negativity are known. In this short paper, we use the minimal value of the Wigner function as such a measure, because it is simple, while providing closed analytical equations. At the same time, in the case of Wigner functions with several local minima, this measure gives only semi-qualitative estimates of the negativity. The detailed analysis based on the volume of the negative part of the Wigner function \cite{Kenfack_JOptB._6_396_2004} will be done in our next paper \cite{Wigner2022}.

We assume here that the Wigner function minimum corresponds to $\alpha_m=0$. If this is not the case, then the quantum state could be adjusted by applying the displacement operator $\hat{D}(-\alpha_m) = e^{\sqrt{\eta}(a_m^*\hat{a} - \alpha_m\hat{a}^\dag)}$, which also takes into account the drift of the minimum due to dissipation.

It is shown in App.\,\ref{app:W0}, that the value of a lossy Wigner function at $\alpha=0$  can be presented as follows:
\begin{multline}\label{W_0}
  W_0(\eta) \equiv W(0,0,\eta) = \frac{2}{\pi}\mean{(1 - 2\eta)^{\hat{n}}} \\
  = \frac{2}{\pi}\sum_{n=0}^\infty(1-2\eta)^n\rho_{nn} \,,
\end{multline}
where $\rho_{nn}=\bra{n}\hat{\rho}\ket{n}$ are the diagonal values of the corresponding density matrix in the number of quanta $\hat{n}=\hat{a}^\dag\hat{a}$ representation. Evidently, to obtain the minimal possible value of $W_0$, all even coefficients $\rho_{nn}$ should be equal to zero:
\begin{equation}\label{even_func}
  \forall n=2k,\ k\ge0:\ \rho_{nn} = 0 \,.
\end{equation}
We assume that the quantum states which we consider here satisfy this condition. In this case,
\begin{equation}\label{W_0_even}
  W_0(\eta) = -\frac{2}{\pi}\mean{(2\eta-1)^{\hat{n}}} \,.
\end{equation}

In the case of small losses, $1-\eta\ll1$, $W_0$ can be approximated as follows:
\begin{equation}\label{W_0_app}
  W_0(\eta) \approx -\frac{2}{\pi}\mean{e^{-2(1-\eta)\hat{n}}}
  = -\frac{2}{\pi}\sum_{n=0}^\infty e^{-2(1-\eta)n}\rho_{nn} \,.
\end{equation}
This form suggests that the negativity actually depends on the product $(1-\eta)n_c$, where $n_c$ is some characteristic number of quanta depending on the probability distribution $\{\rho_{nn}\}$. In particular, if this distribution is well concentrated around its mean value $\bar{n}$, with the valiance $\delta n \ll \bar{n}$, then evidently $n_c=\bar{n}$ and
\begin{equation}\label{W0_asy}
  W_0(\eta) \approx -\frac{2}{\pi}e^{-\varepsilon} \,,
\end{equation}
where
\begin{equation}
  \varepsilon = (1-\eta)\bar{n}
\end{equation}
is the mean number of the lost quanta.

\begin{figure}
  \includegraphics{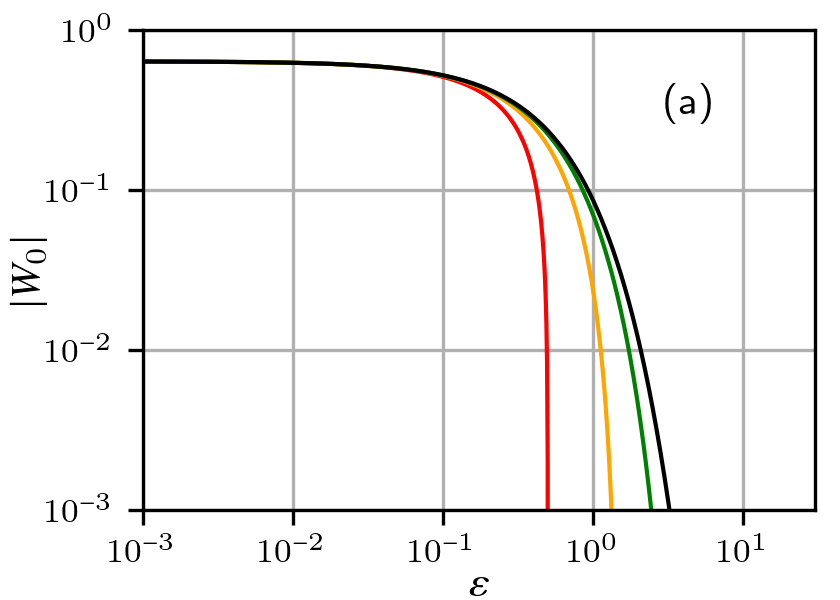} \\
  \includegraphics{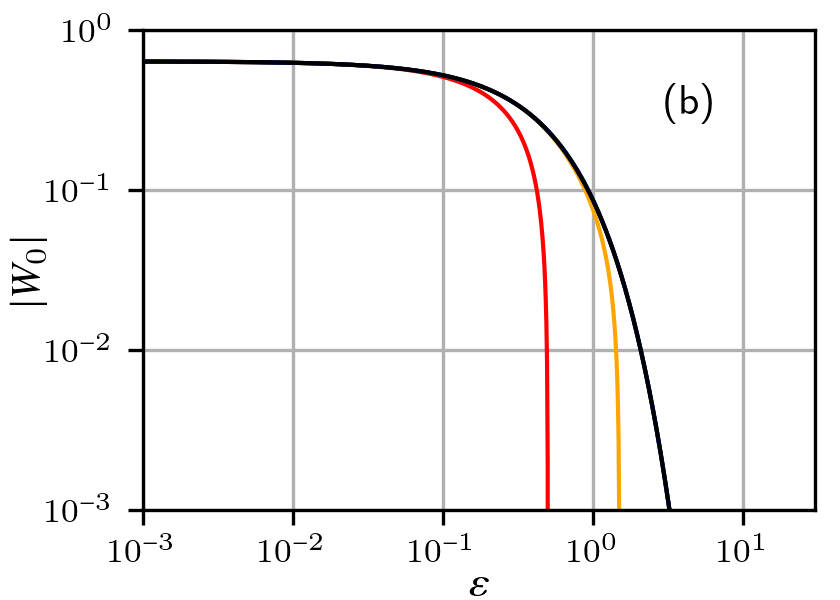} \\
  \includegraphics{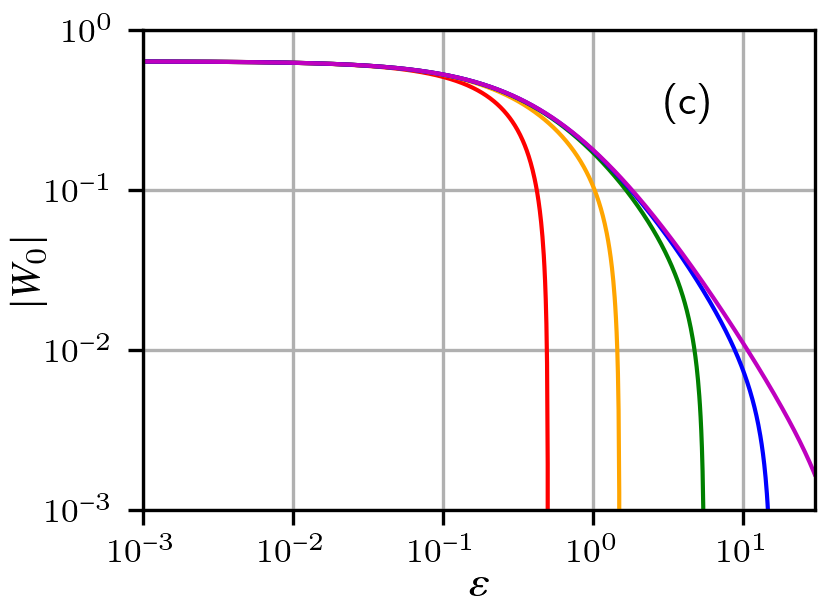}
  \caption{Plots of the absolute values of the Wigner functions minimum as functions of the mean number of lost quanta $\varepsilon$, for the mean number of quanta $\mean{n}=1$ ({\color{red}red}), $3$ ({\color{Orange3}orange}), $11$ ({\color{Green4}green}), 31 ({\color{blue}blue}), $101$ ({\color{Magenta3}magenta}); black lines: the asymptotics \eqref{W0_asy}. (a): Fock state \eqref{W0_Fock}; (b): odd Schrodinger cat state \eqref{W0_odd_cat}; (c): squeezed single photon state \eqref{W0_S1})}\label{fig:plots}
\end{figure}

Consider three characteristic examples, satisfying the condition \eqref{even_func} and being important from the practical point of view.

(i) Fock state $\ket{n}$ with the odd number of quanta $n=2k+1$. In this case,
\begin{equation}\label{W0_Fock}
  W_0(\eta) = -\frac{2}{\pi}(2\eta-1)^n
  = -\frac{2}{\pi}\biggl(1 - \frac{2\varepsilon}{n}\biggr)^n,
\end{equation}

(ii) ``Odd'' Shrodinger cat state \cite{Dodonov_Physica_72_597_1974}
\begin{equation}\label{psi_OC}
  \ket{\tilde{\alpha}_0}
    = \frac{\ket{\alpha_0} - \ket{-\alpha_0}}{[2(1 - e^{-2\alpha_0^2})]^{1/2}}
\end{equation}
(without loss of generality, we assume that $\Im\alpha=0$). It follows from Eqs.\,(\ref{C_eta_1}, \ref{W_eta}, \ref{psi_OC}) that
\begin{equation}\label{W0_odd_cat}
  W_0(\eta)
  = -\frac{2}{\pi}\frac{e^{-2(1-\eta)\alpha_0^2} - e^{-2\eta\alpha_0^2}}{1 - e^{-2\alpha^2}}
\end{equation}
and the mean number of quanta is equal to
\begin{equation}
  \mean{n} = \alpha_0^2\frac{1 + e^{-\alpha_0^2}}{1 - e^{-\alpha_0^2}} \,.
\end{equation}

(iii) Squeezed single quantum state $\hat{S}\ket{1}$, where
$\hat{S} = e^{\frac{r}{2}(\hat{a}^\dag{}^2 - \hat{a}^2)}$
is the squeeze operator and $r$ is the squeeze factor \cite{Lvovsky_RMP_81_299_2009}. In this case (see \eg Ref.\,\cite{17a1KnSpChKh}),
\begin{equation}\label{W0_S1}
  W_0(\eta) = \frac{2}{\pi}\frac{1-2\eta}{(1 + 4\eta(1-\eta)\sinh^2r)^{3/2}} \,.
\end{equation}
and
\begin{equation}
  \mean{n} = 1 + 3\sinh^2r \,.
\end{equation}

In Fig.\,\ref{fig:plots}, the absolute values of the Wigner functions (\ref{W0_Fock}, \ref{W0_odd_cat}, \ref{W0_S1}) are plotted as functions of $\varepsilon$. In the first two cases, the asymptotics \eqref{W0_asy} is also plotted. It can be seen from these plots that for the first two examples, $W_0$ converges quickly to the asymptotic value \eqref{W0_asy}. This is not the case for the third one. The reason for this is evident. The number of quanta distribution of the squeezed single photons state is a very broad one, with $\delta n\sim \bar{n}$ and with a large contribution from the Fock states with small values of $n$. Therefore, the asymptotics \eqref{W_0_even} can not be used for this quantum state.

\paragraph{Conclusion.}

We showed that the condition \eqref{one-half} is not only the necessary \cite{Leonhardt_PRL_72_4086_1994}, but also the sufficient one for the negativity of Wigner function of any pure non-Gaussian state. This means, that in the strict mathematical sense, the negativity always persists for all values of the quantum efficiency $\eta$ down to $1/2$. We explored also the dependence of the Wigner function negativity on $\eta$, using the minimal value of the Wigner function as a simple semi-qualitative measure of negativity. We showed, that in the case of bright (multi-photon) quantum states, that dependence is very nonlinear. After the loss of several photons, which corresponds to $1-\eta \gtrsim 1/\mean{n}$, it could drop by orders of magnitude, which could make its experimental detection problematic.

\acknowledgments

The work of F.Y.K. was supported by the Russian Science Foundation (project 20-12-00344).

\appendix

\section{Proof of Eq.\,\eqref{W_0}}\label{app:W0}

For the case of $s=0$, Eq.\,\eqref{C_eta_1} can be presented as follows:
\begin{equation}
  C(z,s,\eta) = \Tr(\hat{\rho}\hat{K}) \,,
\end{equation}
where
\begin{multline}
  \hat{K}(z) = e^{i\sqrt{\eta}z\hat{a}^\dag}e^{i\sqrt{\eta}z^*\hat{a}}e^{-|z|^2/2} \\
  = \sum_{k,l=1}^\infty\frac{(i\sqrt{\eta}\hat{z})^k(i\sqrt{\eta}z)^l}{k!l!}
      \hat{a}^\dag{}^k\hat{a}^le^{-|z|^2/2} \,.
\end{multline}
Inverse Fourier transform of this operator:
\begin{multline}
  \hat{G}(\alpha) = \intinfty\hat{K}(z)e^{-i(z^*\alpha + z\alpha^*)}\frac{d^2z}{\pi^2} \\
  = \frac{2}{\pi}\sum_{k,l=1}^\infty\frac{(-\sqrt{\eta})^{k+l}\hat{a}^\dag{}^k\hat{a}^l}{k!l!}
      \frac{\partial^{k+l}e^{-2|\alpha|^2}}{\partial\alpha^k\partial\alpha^*{}^l} \,.
\end{multline}
Its value at $\alpha=0$:
\begin{equation}
  \hat{G}(0) = \frac{2}{\pi}\sum_{k=0}^\infty\frac{(-2\eta)^k\hat{a}^\dag{}^k\hat{a}^k}{k!}
  \,.
\end{equation}
Note that $[\hat{G}(0),\hat{n}]=0$. Therefore,
\begin{multline}
  W(0,0,\eta) = \Tr[\hat{\rho}\hat{G}(0)]
  = \frac{2}{\pi}\sum_{k,n=0}^\infty\frac{(-2\eta)^k}{k!}\rho_{nn}
      \bra{n}\hat{a}^\dag{}^k\hat{a}^l\ket{n} \\
  = \frac{2}{\pi}\sum_{n=0}^\infty\rho_{nn}\sum_{k=0}^n\frac{n!}{k!(n-k)!}(-2\eta)^k
  = \frac{2}{\pi}\sum_{n=0}^\infty\rho_{nn}(1-2\eta)^n \,.
\end{multline}


\begin{thebibliography}{22}%
\makeatletter
\providecommand \@ifxundefined [1]{%
 \@ifx{#1\undefined}
}%
\providecommand \@ifnum [1]{%
 \ifnum #1\expandafter \@firstoftwo
 \else \expandafter \@secondoftwo
 \fi
}%
\providecommand \@ifx [1]{%
 \ifx #1\expandafter \@firstoftwo
 \else \expandafter \@secondoftwo
 \fi
}%
\providecommand \natexlab [1]{#1}%
\providecommand \enquote  [1]{``#1''}%
\providecommand \bibnamefont  [1]{#1}%
\providecommand \bibfnamefont [1]{#1}%
\providecommand \citenamefont [1]{#1}%
\providecommand \href@noop [0]{\@secondoftwo}%
\providecommand \href [0]{\begingroup \@sanitize@url \@href}%
\providecommand \@href[1]{\@@startlink{#1}\@@href}%
\providecommand \@@href[1]{\endgroup#1\@@endlink}%
\providecommand \@sanitize@url [0]{\catcode `\\12\catcode `\$12\catcode
  `\&12\catcode `\#12\catcode `\^12\catcode `\_12\catcode `\%12\relax}%
\providecommand \@@startlink[1]{}%
\providecommand \@@endlink[0]{}%
\providecommand \url  [0]{\begingroup\@sanitize@url \@url }%
\providecommand \@url [1]{\endgroup\@href {#1}{\urlprefix }}%
\providecommand \urlprefix  [0]{URL }%
\providecommand \Eprint [0]{\href }%
\providecommand \doibase [0]{https://doi.org/}%
\providecommand \selectlanguage [0]{\@gobble}%
\providecommand \bibinfo  [0]{\@secondoftwo}%
\providecommand \bibfield  [0]{\@secondoftwo}%
\providecommand \translation [1]{[#1]}%
\providecommand \BibitemOpen [0]{}%
\providecommand \bibitemStop [0]{}%
\providecommand \bibitemNoStop [0]{.\EOS\space}%
\providecommand \EOS [0]{\spacefactor3000\relax}%
\providecommand \BibitemShut  [1]{\csname bibitem#1\endcsname}%
\let\auto@bib@innerbib\@empty
\bibitem [{\citenamefont {Braunstein}\ and\ \citenamefont {van
  Loock}(2005)}]{Braunstein_RMP_77_513_2005}%
  \BibitemOpen
  \bibfield  {author} {\bibinfo {author} {\bibfnamefont {S.~L.}\ \bibnamefont
  {Braunstein}}\ and\ \bibinfo {author} {\bibfnamefont {P.}~\bibnamefont {van
  Loock}},\ }\href {https://doi.org/10.1103/RevModPhys.77.513} {\bibfield
  {journal} {\bibinfo  {journal} {Rev. Mod. Phys.}\ }\textbf {\bibinfo {volume}
  {77}},\ \bibinfo {pages} {513} (\bibinfo {year} {2005})}\BibitemShut
  {NoStop}%
\bibitem [{\citenamefont {Cahill}\ and\ \citenamefont
  {Glauber}(1969)}]{Cahill_PR_177_1882_1969}%
  \BibitemOpen
  \bibfield  {author} {\bibinfo {author} {\bibfnamefont {K.~E.}\ \bibnamefont
  {Cahill}}\ and\ \bibinfo {author} {\bibfnamefont {R.~J.}\ \bibnamefont
  {Glauber}},\ }\href {https://doi.org/10.1103/PhysRev.177.1882} {\bibfield
  {journal} {\bibinfo  {journal} {Phys. Rev.}\ }\textbf {\bibinfo {volume}
  {177}},\ \bibinfo {pages} {1882} (\bibinfo {year} {1969})}\BibitemShut
  {NoStop}%
\bibitem [{\citenamefont {{W. Schleich}}(2001)}]{Schleich2001}%
  \BibitemOpen
  \bibfield  {author} {\bibinfo {author} {\bibnamefont {{W. Schleich}}},\
  }\href@noop {} {\emph {\bibinfo {title} {Quantum Optics in Phase Space}}}\
  (\bibinfo  {publisher} {WILEY-VCH, Berlin},\ \bibinfo {year} {2001})\ p.\
  \bibinfo {pages} {695}\BibitemShut {NoStop}%
\bibitem [{\citenamefont {Husimi}(1940)}]{Husimi_PPMSJ3_22-264_1940}%
  \BibitemOpen
  \bibfield  {author} {\bibinfo {author} {\bibfnamefont {K.}~\bibnamefont
  {Husimi}},\ }\href {https://doi.org/10.11429/ppmsj1919.22.4_264} {\bibfield
  {journal} {\bibinfo  {journal} {Proceedings of the Physico-Mathematical
  Society of Japan. 3rd Series}\ }\textbf {\bibinfo {volume} {22}},\ \bibinfo
  {pages} {264} (\bibinfo {year} {1940})}\BibitemShut {NoStop}%
\bibitem [{\citenamefont {Wigner}(1932)}]{Wigner_PR_40_749_1932}%
  \BibitemOpen
  \bibfield  {author} {\bibinfo {author} {\bibfnamefont {E.}~\bibnamefont
  {Wigner}},\ }\href {https://doi.org/10.1103/PhysRev.40.749} {\bibfield
  {journal} {\bibinfo  {journal} {Phys. Rev.}\ }\textbf {\bibinfo {volume}
  {40}},\ \bibinfo {pages} {749} (\bibinfo {year} {1932})}\BibitemShut
  {NoStop}%
\bibitem [{\citenamefont {Glauber}(1963)}]{Glauber_PR_131_2766_1963}%
  \BibitemOpen
  \bibfield  {author} {\bibinfo {author} {\bibfnamefont {R.~J.}\ \bibnamefont
  {Glauber}},\ }\href {https://doi.org/10.1103/PhysRev.131.2766} {\bibfield
  {journal} {\bibinfo  {journal} {Phys. Rev.}\ }\textbf {\bibinfo {volume}
  {131}},\ \bibinfo {pages} {2766} (\bibinfo {year} {1963})}\BibitemShut
  {NoStop}%
\bibitem [{\citenamefont {Vogel}\ and\ \citenamefont
  {Risken}(1989)}]{Vogel_PRA_40_2847_1989}%
  \BibitemOpen
  \bibfield  {author} {\bibinfo {author} {\bibfnamefont {K.}~\bibnamefont
  {Vogel}}\ and\ \bibinfo {author} {\bibfnamefont {H.}~\bibnamefont {Risken}},\
  }\href {https://doi.org/10.1103/PhysRevA.40.2847} {\bibfield  {journal}
  {\bibinfo  {journal} {Phys. Rev. A}\ }\textbf {\bibinfo {volume} {40}},\
  \bibinfo {pages} {2847} (\bibinfo {year} {1989})}\BibitemShut {NoStop}%
\bibitem [{\citenamefont {Hudson}(1974)}]{Hudson_RMP_6_249_1974}%
  \BibitemOpen
  \bibfield  {author} {\bibinfo {author} {\bibfnamefont {R.}~\bibnamefont
  {Hudson}},\ }\href
  {https://doi.org/https://doi.org/10.1016/0034-4877(74)90007-X} {\bibfield
  {journal} {\bibinfo  {journal} {Reports on Mathematical Physics}\ }\textbf
  {\bibinfo {volume} {6}},\ \bibinfo {pages} {249} (\bibinfo {year}
  {1974})}\BibitemShut {NoStop}%
\bibitem [{\citenamefont {Marshall}\ \emph {et~al.}(2003)\citenamefont
  {Marshall}, \citenamefont {Simon}, \citenamefont {Penrose},\ and\
  \citenamefont {Bouwmeester}}]{Marshall_PRL_91_130401_2003}%
  \BibitemOpen
  \bibfield  {author} {\bibinfo {author} {\bibfnamefont {W.}~\bibnamefont
  {Marshall}}, \bibinfo {author} {\bibfnamefont {C.}~\bibnamefont {Simon}},
  \bibinfo {author} {\bibfnamefont {R.}~\bibnamefont {Penrose}},\ and\ \bibinfo
  {author} {\bibfnamefont {D.}~\bibnamefont {Bouwmeester}},\ }\href
  {https://doi.org/10.1103/PhysRevLett.91.130401} {\bibfield  {journal}
  {\bibinfo  {journal} {Phys. Rev. Lett.}\ }\textbf {\bibinfo {volume} {91}},\
  \bibinfo {pages} {130401} (\bibinfo {year} {2003})}\BibitemShut {NoStop}%
\bibitem [{\citenamefont {Romero-Isart}\ \emph {et~al.}(2010)\citenamefont
  {Romero-Isart}, \citenamefont {Juan}, \citenamefont {Quidant},\ and\
  \citenamefont {Cirac}}]{Romero-Isart_NJP_12_033015_2010}%
  \BibitemOpen
  \bibfield  {author} {\bibinfo {author} {\bibfnamefont {O.}~\bibnamefont
  {Romero-Isart}}, \bibinfo {author} {\bibfnamefont {M.~L.}\ \bibnamefont
  {Juan}}, \bibinfo {author} {\bibfnamefont {R.}~\bibnamefont {Quidant}},\ and\
  \bibinfo {author} {\bibfnamefont {J.~I.}\ \bibnamefont {Cirac}},\ }\href
  {http://stacks.iop.org/1367-2630/12/i=3/a=033015} {\bibfield  {journal}
  {\bibinfo  {journal} {New Journal of Physics}\ }\textbf {\bibinfo {volume}
  {12}},\ \bibinfo {pages} {033015} (\bibinfo {year} {2010})}\BibitemShut
  {NoStop}%
\bibitem [{\citenamefont {Khalili}\ \emph {et~al.}(2010)\citenamefont
  {Khalili}, \citenamefont {Danilishin}, \citenamefont {Miao}, \citenamefont
  {M{\"u}ller-Ebhardt}, \citenamefont {Yang},\ and\ \citenamefont
  {Chen}}]{10a1KhDaMiMuYaCh}%
  \BibitemOpen
  \bibfield  {author} {\bibinfo {author} {\bibfnamefont {F.}~\bibnamefont
  {Khalili}}, \bibinfo {author} {\bibfnamefont {S.}~\bibnamefont {Danilishin}},
  \bibinfo {author} {\bibfnamefont {H.}~\bibnamefont {Miao}}, \bibinfo {author}
  {\bibfnamefont {H.}~\bibnamefont {M{\"u}ller-Ebhardt}}, \bibinfo {author}
  {\bibfnamefont {H.}~\bibnamefont {Yang}},\ and\ \bibinfo {author}
  {\bibfnamefont {Y.}~\bibnamefont {Chen}},\ }\href
  {https://doi.org/10.1103/PhysRevLett.105.070403} {\bibfield  {journal}
  {\bibinfo  {journal} {Phys. Rev. Lett.}\ }\textbf {\bibinfo {volume} {105}},\
  \bibinfo {pages} {070403} (\bibinfo {year} {2010})}\BibitemShut {NoStop}%
\bibitem [{\citenamefont {Lvovsky}\ \emph {et~al.}(2020)\citenamefont
  {Lvovsky}, \citenamefont {Grangier}, \citenamefont {Ourjoumtsev},
  \citenamefont {Parigi}, \citenamefont {Sasaki},\ and\ \citenamefont
  {Tualle-Brouri}}]{Lvovsky_2006_16985}%
  \BibitemOpen
  \bibfield  {author} {\bibinfo {author} {\bibfnamefont {A.~I.}\ \bibnamefont
  {Lvovsky}}, \bibinfo {author} {\bibfnamefont {P.}~\bibnamefont {Grangier}},
  \bibinfo {author} {\bibfnamefont {A.}~\bibnamefont {Ourjoumtsev}}, \bibinfo
  {author} {\bibfnamefont {V.}~\bibnamefont {Parigi}}, \bibinfo {author}
  {\bibfnamefont {M.}~\bibnamefont {Sasaki}},\ and\ \bibinfo {author}
  {\bibfnamefont {R.}~\bibnamefont {Tualle-Brouri}},\ }\href@noop {} {\bibinfo
  {title} {Production and applications of non-gaussian quantum states of
  light}} (\bibinfo {year} {2020}),\ \Eprint {https://arxiv.org/abs/2006.16985}
  {arXiv:2006.16985 [quant-ph]} \BibitemShut {NoStop}%
\bibitem [{\citenamefont {Walschaers}(2021)}]{Walschaers_PRXQ_2_030204_2021}%
  \BibitemOpen
  \bibfield  {author} {\bibinfo {author} {\bibfnamefont {M.}~\bibnamefont
  {Walschaers}},\ }\href {https://doi.org/10.1103/PRXQuantum.2.030204}
  {\bibfield  {journal} {\bibinfo  {journal} {PRX Quantum}\ }\textbf {\bibinfo
  {volume} {2}},\ \bibinfo {pages} {030204} (\bibinfo {year}
  {2021})}\BibitemShut {NoStop}%
\bibitem [{\citenamefont {Bell}(1987)}]{Bell_1987}%
  \BibitemOpen
  \bibfield  {author} {\bibinfo {author} {\bibfnamefont {J.~S.}\ \bibnamefont
  {Bell}},\ }\href@noop {} {\emph {\bibinfo {title} {Speakable and Unspeakable
  in Quantum Mechanics}}}\ (\bibinfo  {publisher} {Cambridge University Press,
  Cambridge, UK},\ \bibinfo {year} {1987})\ p.\ \bibinfo {pages}
  {212}\BibitemShut {NoStop}%
\bibitem [{\citenamefont {Mari}\ and\ \citenamefont
  {Eisert}(2012)}]{Mari_PRL_109_230503_2012}%
  \BibitemOpen
  \bibfield  {author} {\bibinfo {author} {\bibfnamefont {A.}~\bibnamefont
  {Mari}}\ and\ \bibinfo {author} {\bibfnamefont {J.}~\bibnamefont {Eisert}},\
  }\href {https://doi.org/10.1103/PhysRevLett.109.230503} {\bibfield  {journal}
  {\bibinfo  {journal} {Phys. Rev. Lett.}\ }\textbf {\bibinfo {volume} {109}},\
  \bibinfo {pages} {230503} (\bibinfo {year} {2012})}\BibitemShut {NoStop}%
\bibitem [{\citenamefont {Zurek}(2003)}]{Zurek_RMP_75_715_2003}%
  \BibitemOpen
  \bibfield  {author} {\bibinfo {author} {\bibfnamefont {W.~H.}\ \bibnamefont
  {Zurek}},\ }\href {https://doi.org/10.1103/RevModPhys.75.715} {\bibfield
  {journal} {\bibinfo  {journal} {Rev. Mod. Phys.}\ }\textbf {\bibinfo {volume}
  {75}},\ \bibinfo {pages} {715} (\bibinfo {year} {2003})}\BibitemShut
  {NoStop}%
\bibitem [{\citenamefont {Leonhardt}\ and\ \citenamefont
  {Paul}(1994)}]{Leonhardt_PRL_72_4086_1994}%
  \BibitemOpen
  \bibfield  {author} {\bibinfo {author} {\bibfnamefont {U.}~\bibnamefont
  {Leonhardt}}\ and\ \bibinfo {author} {\bibfnamefont {H.}~\bibnamefont
  {Paul}},\ }\href {https://doi.org/10.1103/PhysRevLett.72.4086} {\bibfield
  {journal} {\bibinfo  {journal} {Phys. Rev. Lett.}\ }\textbf {\bibinfo
  {volume} {72}},\ \bibinfo {pages} {4086} (\bibinfo {year}
  {1994})}\BibitemShut {NoStop}%
\bibitem [{\citenamefont {Kenfack}\ and\ \citenamefont
  {Zyczkowski}(2004)}]{Kenfack_JOptB._6_396_2004}%
  \BibitemOpen
  \bibfield  {author} {\bibinfo {author} {\bibfnamefont {A.}~\bibnamefont
  {Kenfack}}\ and\ \bibinfo {author} {\bibfnamefont {K.}~\bibnamefont
  {Zyczkowski}},\ }\href {https://doi.org/10.1088/1464-4266/6/10/003} {\bibfield
   {journal} {\bibinfo  {journal} {Journal of Optics B: Quantum and
  Semiclassical Optics}\ }\textbf {\bibinfo {volume} {6}},\ \bibinfo {pages}
  {396} (\bibinfo {year} {2004})}\BibitemShut {NoStop}%
\bibitem [{\citenamefont {Nugmanov}\ \emph {et~al.}(2022)\citenamefont
  {Nugmanov}, \citenamefont {Zunikov},\ and\ \citenamefont
  {Khalili}}]{Wigner2022}%
  \BibitemOpen
  \bibfield  {author} {\bibinfo {author} {\bibfnamefont {B.}~\bibnamefont
  {Nugmanov}}, \bibinfo {author} {\bibfnamefont {N.}~\bibnamefont {Zunikov}},\
  and\ \bibinfo {author} {\bibfnamefont {F.}~\bibnamefont {Khalili}},\
  }\href@noop {} {} (\bibinfo {year} {2022}),\ \bibinfo {note} {{\it paper in
  prepration}}\BibitemShut {NoStop}%
\bibitem [{\citenamefont {Dodonov}\ \emph {et~al.}(1974)\citenamefont
  {Dodonov}, \citenamefont {Malkin},\ and\ \citenamefont
  {Man'ko}}]{Dodonov_Physica_72_597_1974}%
  \BibitemOpen
  \bibfield  {author} {\bibinfo {author} {\bibfnamefont {V.}~\bibnamefont
  {Dodonov}}, \bibinfo {author} {\bibfnamefont {I.}~\bibnamefont {Malkin}},\
  and\ \bibinfo {author} {\bibfnamefont {V.}~\bibnamefont {Man'ko}},\ }\href
  {https://doi.org/https://doi.org/10.1016/0031-8914(74)90215-8} {\bibfield
  {journal} {\bibinfo  {journal} {Physica}\ }\textbf {\bibinfo {volume} {72}},\
  \bibinfo {pages} {597} (\bibinfo {year} {1974})}\BibitemShut {NoStop}%
\bibitem [{\citenamefont {Lvovsky}\ and\ \citenamefont
  {Raymer}(2009)}]{Lvovsky_RMP_81_299_2009}%
  \BibitemOpen
  \bibfield  {author} {\bibinfo {author} {\bibfnamefont {A.~I.}\ \bibnamefont
  {Lvovsky}}\ and\ \bibinfo {author} {\bibfnamefont {M.~G.}\ \bibnamefont
  {Raymer}},\ }\href {https://doi.org/10.1103/RevModPhys.81.299} {\bibfield
  {journal} {\bibinfo  {journal} {Rev. Mod. Phys.}\ }\textbf {\bibinfo {volume}
  {81}},\ \bibinfo {pages} {299} (\bibinfo {year} {2009})}\BibitemShut
  {NoStop}%
\bibitem [{\citenamefont {Knyazev}\ \emph {et~al.}(2018)\citenamefont
  {Knyazev}, \citenamefont {Spasibko}, \citenamefont {Khalili},\ and\
  \citenamefont {Chekhova}}]{17a1KnSpChKh}%
  \BibitemOpen
  \bibfield  {author} {\bibinfo {author} {\bibfnamefont {E.}~\bibnamefont
  {Knyazev}}, \bibinfo {author} {\bibfnamefont {K.}~\bibnamefont {Spasibko}},
  \bibinfo {author} {\bibfnamefont {F.}~\bibnamefont {Khalili}},\ and\ \bibinfo
  {author} {\bibfnamefont {M.}~\bibnamefont {Chekhova}},\ }\href
  {https://doi.org/https://doi.org/10.1088/1367-2630/aa99b4} {\bibfield
  {journal} {\bibinfo  {journal} {New Journal of Physics}\ }\textbf {\bibinfo
  {volume} {20}},\ \bibinfo {pages} {013005} (\bibinfo {year}
  {2018})}\BibitemShut {NoStop}%
\end{thebibliography}

%

\end{document}